\documentclass[12pt]{article}

\usepackage{amsmath,srcltx}
\usepackage{graphicx}  
\usepackage{epsfig}  
\usepackage{amssymb}
\usepackage{latexsym}
\usepackage{axodraw}
\makeatletter 
\@addtoreset{equation}{section}  
\makeatother 
\renewcommand{\theequation}{\thesection.\arabic{equation}} 

\newcommand{\Hil}{\mathcal{H}}

\def\ii{\'\i}
\def\ao{\~ao}

\def\ii{\'\i}
\def\ao{\~ao}

\def\ftoday{{\sl {Le \number\day \space\ifcase\month 
\or janvier\or f\'evrier\or mars\or avril\or mai
\or juin\or juillet\or ao\^ut\or septembre\or octobre
\or novembre \or d\'ecembre\fi\space \number\year}}}    
\def\ptoday{{\sl {\number\day \space de\space \ifcase\month 
\or janeiro\or fevereiro\or mar{\c c}o\or abril\or maio
\or junho\or julho\or agosto\or setembro\or outubro
\or novembro \or dezembro\fi\space de\space \number\year}}}    
\def\gtoday{{\sl {Den \number\day. \ifcase\month 
\or Januar\or Februar\or M\"arz\or April\or Mai
\or Juni\or Juli\or August\or September\or Oktober
\or November \or Dezember\fi\space \number\year}}}    
\def\today{{\sl {\ifcase\month
\or January\or February\or March\or April\or May
\or June\or July\or August\or September\or October
\or November \or December\fi \space\number\day,\space 
                                            \number\year}}}
\newcommand{\journa}[4]{{\em #1~}#2\,(#3)\,#4}

\newcommand{\ijmp}{\journa {Int. J. Mod. Phys.}}

\newcommand{\pr}{\journa {Phys. Rev.}}

\newcommand{\prl}{\journa {Phys. Rev. Lett.}}
\newcommand{\jmp}{\journa {J. Math. Phys.}}

\newcommand{\cmp}{\journa {Commun. Math. Phys.}}
\newcommand{\lmp}{\journa {Lett. Math. Phys.}}

\newcommand{\cqg}{\journa {Class. Quantum Grav.}}

\newcommand{\np}{\journa {Nucl. Phys.}}
\newcommand{\pl}{\journa {Phys. Lett.}}

\newcommand{\prep}{\journa {Phys. Rep.}}

\newcommand{\annp}{\journa {Annals Phys.}}

\renewcommand{\a}{\alpha}
\renewcommand{\b}{\beta}
\newcommand{\g}{\gamma}           \newcommand{\GA}{\Gamma}
\renewcommand{\d}{\delta}         
\newcommand{\varep}{\varepsilon}

\newcommand{\p}{\psi}             
           \renewcommand{\S}{\Sigma}
\newcommand{\f}{{\phi}}           \newcommand{\F}{{\Phi}}

\newcommand{\HH}{{\cal H}}

\newcommand{\LL}{{\cal L}}

\newcommand{\sla}{\raise.15ex\hbox{$/$}\kern -.57em} 
\newcommand{\Sla}{\raise.15ex\hbox{$/$}\kern -.70em}

\newcommand{\lp}{\left(}\newcommand{\rp}{\right)}

\newcommand{\lac}{\left\{}\newcommand{\rac}{\right\}}

\newcommand{\complex}{{\kern .1em {\raise .47ex
\hbox {$\scriptscriptstyle |$}}
    \kern -.4em {\rm C}}}
\newcommand{\real}{{{\rm I} \kern -.19em {\rm R}}}
\newcommand{\rational}{{\kern .1em {\raise .47ex
\hbox{$\scripscriptstyle |$}}
    \kern -.35em {\rm Q}}}
\renewcommand{\natural}{{\vrule height 1.6ex width
.05em depth 0ex \kern -.35em {\rm N}}}

\newcommand{\Tr}{{\rm {Tr} \,}}
\newcommand{\half}{\frac{1}{2}}
\newcommand{\pa}{\partial}

\newcommand{\dsum}[2]{\displaystyle{\sum_{#1}^{#2}}}   
\newcommand{\dint}{\displaystyle{\int}}

\newcommand{\twiddle}{\lower.9ex\rlap{$\kern -.1em\scriptstyle\sim$}}

\newcommand{\bra}[1]{\left\langle {#1}\right|}
\newcommand{\ket}[1]{\left| {#1}\right\rangle}

\newcommand{\vev}[1]{\left\langle {#1}\right\rangle}
\newcommand{\equ}[1]{(\ref{#1})}
\newcommand{\eq}{\begin{equation}}
\newcommand{\eqn}[1]{\label{#1}\end{equation}}
\newcommand{\eea}{\end{eqnarray}}
\newcommand{\eqa}{\begin{eqnarray}}
\newcommand{\eqan}[1]{\label{#1}\end{eqnarray}}
\newcommand{\ba}{\begin{array}}
\newcommand{\ea}{\end{array}}
\newcommand{\eqac}{\begin{equation}\begin{array}{rcl}}
\newcommand{\eqacn}[1]{\end{array}\label{#1}\end{equation}}

\newcommand{\bz}{\begin{enumerate}}
\newcommand{\ez}{\end{enumerate}}

\title{The Hilbert space of Chern-Simons theory on the cylinder. 
A Loop Quantum Gravity approach}
\author{Clisthenis P. Constantinidis\footnote{Work supported
   in part by the Conselho Nacional
   de Desenvolvimento Cient\'{\i}fico e
   Tecnol\'{o}gico -- CNPq (Brazil) and 
by the PRONEX project No. 35885149/2006 from FAPES -- CNPq (Brazil).}$^{\ ,1}$,
Gabriel Luchini$^{*,2}$ \\
and Olivier Piguet$^{*,1}$
}

\begin{document}
\maketitle

\begin{center} 

$^1$Departamento de F\ii sica, Universidade Federal do Espirito Santo (UFES)\\
 Vit\'oria, ES, Brazil\\
$^2$Instituto de F\ii sica de S\ao\ Carlos (IFSC),
Universidade de S\ao\ Paulo (USP)\\
S\ao\ Carlos, SP, Brasil

\vspace{5mm}

{\small\tt E-mails: cpconstantinidis@pq.cnpq.br, 
gabriel.luchini@ursa.ifsc.usp.br,
opiguet@pq.cnpq.br}
\end{center}


\begin{abstract}
As a laboratory for loop quantum gravity, we consider 
the canonical quantization of the three-dimensional 
Chern-Simons theory on a  noncompact space with the topology 
of a cylinder. 
Working within the loop quantization formalism, 
we define at the quantum level the constraints appearing in the 
canonical approach and completely solve them, thus constructing  a gauge 
and diffeomorphism invariant physical Hilbert space
 for the theory.  This space turns out to be infinite dim\-en\-sion\-al, 
but separable. 
\end{abstract}

\section{Introduction}

Chern-Simons (CS) topological 
theory~\cite{schwarz,deser-jackiw,witten2,witten1,topol-phys-rep} 
is one of the simplest field theoretic systems
showing general covariance, i.e. full invariance under space-time
diffeomorphisms, which characterizes it as a background independent theory.
Far simpler than genuine gravitation theory in higher dimensional
space-times and already very well 
studied~\cite{witten2,elitzur,zoo,djt,gmm,bos-nair},
this 3-dimensional topological theory however deserves a
study in  the ``loop quantization'' framework introduced for the
canonical quantization of General Relativity and described in the books
and review  papers~\cite{rovelli,thiemann,ash-lew,han,perez-pedra-azul}, 
where references to
the numerous original papers can be found. As far as the authors know,
the application of this scheme to the CS theory has not yet
appeared in the literature, although it has been successfully applied to various
other topological models of the Schwarz type~\cite{schwarz,topol-phys-rep}, 
such as $BF$ theories~\cite{baez}, 
and to low dimensional gravitation 
theories~\cite{perez,garcia,cons-perez-pig} -- which are also
Schwarz topological theories.

The starting point is the canonical quantization program of Dirac~\cite{dirac,ht} which,
in the case of a generally covariant theory, involves a Hamiltonian purely
made of constraints. One first constructs a kinematical Hilbert space
where the fields are represented by operators, 
and then selects the physical states as the vectors satisfying the
constraints. Since the constraint operators generate the gauge
invariances of the theory, the physical states are in fact the gauge invariant vectors. 

A peculiarity of CS theory is that the space components of the 
gauge connection $A$ form a pair of conjugate 
variables, so that
the  wave functional $\Psi$ in the Schr\"odinger picture is a function of
one of these components, let us say $A_1$. Then $A_2$ is represented by
a functional derivative. On the other hand, in topological theories of the
Schwarz type, diffeomorphism invariance is a simple consequence of gauge
invariance~\cite{witten1}, at least at the classical level. 
One would therefore expect that, applying the Gauss constraint which
ensures spatial gauge invariance, one would automatically ensure
invariance under spatial diffeomorphisms and thus determine the physical
Hilbert space.
We will however see that, due
to the necessity of choosing  a polarization, i.e. choosing
which component of $A$ plays the role of 
a coordinate and which one 
plays the role of a momentum, diffeomorphism invariance is not automatic
and must be implemented at the end as another constraint.

The canonical formalism requires the space-time topology to be that of 
$\real\times\Sigma$ where $\real$ stands for the time dimension and $\Sigma$
for a space slice. In order to proceed with some details a topology  for space
must also be specified. We will choose that of  a noncompact space,
namely of a cylinder: $\S$ $=$ $\real\times S^1$.  As we will see, this
choice leads to an infinite dimensional physical Hilbert space. 
To the best of our knowledge, nonperturbative quantization
in the case of a noncompact space slice has
not yet been considered in the literature, except the case of $\real^2$,
which leads to a 1-dimensional Hilbert space~\cite{djt,gmm}. Apart of
the latters, published results\footnote{ See quoted references, 
and~\cite{topol-phys-rep} for more references.} concern compact closed 
spaces, where Hilbert space is
finite dimensional, as well as spaces with punctures
or with boundary, where exist the local degrees of
freedom of a two-dimensional conformal theory~\cite{witten2,zoo}.

Let us remind that the approach of
the present paper is of the type ``quantize first and then apply the
constraints'' and has been applied to CS theory by various authors,
in particular by~\cite{djt,gmm}. It has to
be contrasted with the approach ``reduce first the classical phase space by
imposing the constraints there, and then quantize'', which has been used
in particular by Witten in his pioneering paper~\cite{witten2}
 (see also~\cite{bos-nair}). Both 
approaches may lead to inequivalent quantum field
theories~\cite{djt}.

We will essentially follow the reference~\cite{djt} for the statement of the
problem and will use some of its results and notations. 
Our own contribution is an explicit
construction of the kinematical and physical Hilbert spaces with a
well defined internal product. We will restrict  ourselves to a compact
semi-simple Lie group of gauge invariance, typically SU(2), in order
to avoid the difficulties which may arise in the noncompact case in the
definition of the internal product~\cite{freidel-livine}.

After briefly recalling in Section \ref{sec-classical} some basic facts on the 
classical CS theory in the canonical framework, we proceed to the
construction of the quantum state space in 
Section \ref{sec-hilbertspace}. Concluding remarks are presented in Section
\ref{conclusions}.


\section{Classical Chern-Simons theory \emph{\`a la} 
Dirac }\label{sec-classical}

Chern-Simons theory, being a fully constrained theory, may conveniently  
be treated using Dirac's method~\cite{dirac,ht,mat}, 
which in turn, arises from the canonical point of view. 
In this Section 
devoted to the classical theory, we follow\footnote{The canonical
formalism for Chern-Simons theory may be found in~\cite{deser-xiang}.}
the reference~\cite{djt}.
The action is given by 
\begin{equation}
\label{action}
 S =-\kappa \int_{\mathcal{M}}d^{3}x\; 
\epsilon^{\mu \nu \rho}\;{\rm Tr}\Big( A_{\mu}\partial_{\nu}A_{\rho}
+\frac{2}{3}A_{\mu}A_{\nu}A_{\rho} \Big)\,.
\end{equation}
The dynamical field $A=A_{\mu}^{I}\tau_{I}dx^{\mu}$ with $\mu=0,1,2$, 
is a Lie algebra valued connection 1-form. 
The gauge Lie group  $G$ will be assumed to be compact and semi-simple.
The generators of the Lie algebra\footnote{In the case of the SU(2) group 
which will be used throughout this paper, one has $\tau_I=-\frac{i}{2}\sigma_I$
(I=1,2,3), and $f_{IJ}{}^K=\epsilon_{IJK}$.} 
satisfy the product 
$[\tau_{I},\tau_{J}]=f_{IJ}{}^K\tau_K$, 
where $f_{IJ}{}^K$ are the algebra's structure constants
 and ${\rm Tr}(\tau_I\tau_J)$ $=$ $-\half\d_{IJ}$. 
The tensor $f_{IJK}$ is completely antisymmetric in its three indices.
The $\kappa$ appearing in the above equation is the coupling constant which  
is well-known to be quantized~\cite{deser-jackiw,witten2}, $\kappa$ $=$ $\frac{n}{4\pi}$, 
due the  gauge invariance of the quantum  path integral.

The model is a generally covariant theory, integration in 
\equ{action} being performed on a ``space-time'' 3-manifold $\mathcal{M}$
without metric structure. Hence there is no \emph{a priori} notion of ``time''.
However, in a canonical approach, a time variable is introduced
through the hypothesis that space-time has the topological structure of
$\Sigma \times \real$, where ``space'' is given by the 2-dimensional 
hypersurface $\Sigma$ and ``time'' by the real line $\real$.
The action then reads
\begin{equation}
\label{action2}
 S =-\kappa \int_{\real}\int_{\Sigma}dt\; 
d^{2}x\;\epsilon^{ab}\;{\rm Tr}\Big( \dot{A}_{a}A_{b}+A_{0}F_{ab}\Big)\,,
\end{equation}
with the spatial curvature given by 
$F^{I}_{ab}=\partial_{a}A^{I}_{b}-\partial_{b}A^{I}_{a}+f_{JK}{}^I
A^{J}_{a}A^{K}_{b}$,
$a,b,\cdots=1,2$.

An analysis according to the Dirac-Bergman's algorithm~\cite{dirac,ht}
leads to a symplectic structure corresponding to the following Dirac
brackets:
\eq\ba{c}
\{A_{a}^{I}(x),A_{b}^{J}(y)\}
=\frac{1}{\kappa}\epsilon_{ab}\delta^{IJ}\delta^{2}(x-y)\,,
\ea\eqn{aa}
and to the Hamiltonian
\begin{equation}
H = G(\varep) := \int_{\Sigma}d^{2}x\;\varep^{I}(x)G^{I}(x)\approx 0\,,
\end{equation}
where $\varep^I$  is an arbitrary test function and 
$G^I$ the Gauss constraint
\begin{equation}
\label{g}
G^{I}:=-\frac{\kappa}{2}\epsilon^{ab}F^{I}_{ab}\approx 0\,.
\end{equation}
We see that this is a completely constrained system and that 
the space components $A_1^I$ and $A_2^I$ form a pair of conjugate
variables.
The Gauss constraint is first class and its Dirac bracket algebra
reproduces the Lie algebra of the gauge group:
\eq
\{G(\varep), G(\varep')\} = G(\varep\times\varep')\,,\quad
(\varep\times\varep')^I:=f_{JK}{}^I\varep^J{\varep'}^K\,.
\eqn{brackets-G-phi}
It generates the space gauge transformations
\eq
\{ A_{a}^{I}(x), G(\varep^{I})\}
=  D_a \varep^{I}(x)
=  \pa_a \varep^{I}(x)  +f_{JK}{}^I A_a^{J}(x) \varep^{K}(x) \,.
\eqn{smeared-H}
 We finally remember that diffeomorphism invariance -- which is explicitly
verified by the original action \equ{action}, can be shown to follow
directly from gauge invariance. In particular, implementation of the
Gauss constraint guaranties invariance under spatial diffeomorphisms. 
Indeed, such a diffeomorphism is given in the infinitesimal form by
the Lie derivative $\LL_\xi$ along a spatial vector field
$\xi=(\xi^z,\xi^\theta)$, and one easily checks that it is equal to a
gauge transformation with parameter $\xi^aA_a^I$,
up to a term proportional to the Gauss constraint:
\eq
\LL_\xi A_a = \xi^b F_{ba} + D_a(\xi^bA_b) \approx D_a(\xi^bA_b) \,.
\eqn{diff-A}

\section{Construction of the Hilbert space}
\label{sec-hilbertspace}

Quantization \emph{\`a la} Dirac is performed in two steps, namely construct first
a kinematical Hilbert space  $\HH_{\rm kin}$  based on the phase space coordinates
provided by the gauge connection, more precisely its space components
$A_a$, and then select the physical states through 
the constraints which, in the present case, are 
given by \equ{g}. 

\subsection{The kinematical Hilbert space}\label{sec-kinspace}

Working in the Schr\"odinger picture, we choose the spatial components
$A^I_\theta$ $:=$ $A^I_1$ as the generalized coordinates and 
$A^I_z$ $:=$ $A^I_2$ as the generalized
momenta obeying, as operators, the canonical commutation relations
\begin{equation}
\label{quantumcom}
[\hat{A}_{\theta}^{I}(x),\hat{A}_{z}^{J}(y)]=\frac{i}{\kappa}\delta^{IJ}\delta^{2}(x-y)
\end{equation}
corresponding to the Dirac bracket relations \equ{aa} of the classical
theory\footnote{We take $\hbar=1$.}. The indices $z$ and $\theta$ refer to a special choice of
coordinates $z\in\real$ and $\theta$ ($0\leq\theta\leq2\pi$) 
adapted to the topological configuration 
we are demanding for
the 2-dimensional  space $\S$, namely that of a cylinder
$\real\times S^1$. States are described by wave functionals 
-- in Dirac's notation: $\Psi[A_{\theta}]=\bra{A_{\theta}}\Psi\rangle$.
The field operators act on these states as
\eq
\bra{A_{\theta}}\hat{A}_{\theta}^{I}(x)\ket{\Psi}
= A_{\theta}^{I}(x)\Psi[A_{\theta}]\,,
\qquad
\bra{A_{\theta}}\hat{A}_{z}^{I}(x)\ket{\Psi}
= \frac{1}{i\kappa}\frac{\delta}{\delta
A_{\theta}^{I}(x)}\Psi[A_{\theta}]\,.
\eqn{ope1}
 This choice of polarization, namely of $A^I_{\theta}$ as the
configuration variables and of ${A}^I_{z}$ as the momentum variables
will turn out to be the more adequate to our aim of describing quantum
states as gauge invariant functions of holonomies.

Up to now everything we have written remains purely formal until we
define an integration measure in configuration space allowing us to
define an internal product between the state vectors $\ket{\Psi}$.
Before doing this, let us already examine the effect of the Gauss
constraint at this formal level. It acts on the 
wave functionals as the operator
\begin{equation}
\hat G{}^I \Psi[A_{\theta}] 
=  i\Bigg( \frac{\partial}{\partial \theta}\frac{\delta}{\delta A_{\theta}^{I}}
+f_{IJ}{}^K A_{\theta}^{J}\frac{\delta}{\delta A_{\theta}^{K}} \Bigg)\Psi[A_{\theta}]+
 \kappa \Bigg( \frac{\partial}{\partial z}A^I_{\theta}
\Bigg)\Psi[A_{\theta}]\,.
\eqn{quantumgauss-op}
A particular solution of the Gauss constraint 
\eq
\hat G{}^I \Psi[A_{\theta}] = 0\,,
\eqn{quantumgauss}
is provided by the phase~\cite{djt}
\eq
\Psi_{\circ}[A_\theta] = e^{2\pi i\alpha_{\circ}}\,,
\eqn{Psi_0} 
with 
\eq
\alpha_{\circ}=4\pi \kappa \int_{\tilde{\S}} d^{3}x\; w(g) 
- \frac{\kappa}{2\pi}\int_\S 
d^{2}x\;{\rm Tr}(A_{\theta}g^{-1}\partial_{z}g)\,,
\eqn{bbbb}
where the first integral is performed on a 3-dimensional manifold
$\tilde{\S}$  whose boundary is space $\S$,
and $g\in G$ is defined in terms of $A_{\theta}$ by
$A_{\theta}=g^{-1}\partial_{\theta}g$. 
The integral of $w$ over $\tilde{\S}$,
\begin{equation}
\label{w}
\int_{\tilde{\S}}d^{3}x\;w:
= \frac{1}{24\pi^{2}}\;\int_{\tilde{\S}}d^{3}x\;\Big(\epsilon^{\mu \nu \rho} 
{\rm Tr}(g^{-1}\partial_{\mu}g\;g^{-1}\partial_{\nu}g\;
g^{-1}\partial_{\rho}g)\Big)\,,
\end{equation}
 called the Wess-Zumino-Witten action, is an integer for the case where $G$ is 
a nonabelian compact group, and consequently,  a singlevalued wave functional  
requires then the quantization of the coupling constant: 
\eq
\kappa=\frac{n}{4\pi}\,,\quad n \in \mathbb{Z}\,.
\eqn{quant-kappa}
Moreover~\cite{djt}, the general solution of \equ{quantumgauss} is given by 
\eq
\Psi[A_{\theta}] = \Psi_{\circ}[A_\theta] \psi^{\rm inv}[A_{\theta}]\,,
\eqn{gsol}
where $\psi^{\rm inv}[A_{\theta}]$  -- the ``reduced functional'' -- is  
``$\theta$-gauge invariant'', i.e., obeys the condition
\eq
i\Bigg( \frac{\partial}{\partial \theta}\frac{\delta}{\delta A_{\theta}^{I}}
+f_{IJ}{}^K A_{\theta}^{J}\frac{\delta}{\delta A_{\theta}^{K}} \Bigg)
\psi^{\rm inv}[A_{\theta}]=0\,,
\eqn{reduced-func}
where the functional derivative operator in the left-hand side is the
first part of the Gauss constraint operator \equ{quantumgauss-op} and generates 
the $\theta$-gauge  transformation
\eq
\d A^I_\theta = D_\theta\varep^I\,,\quad \d A^I_z= f_{JK}{}^I A^I_z \varep^K \,.
\eqn{theta-transf}

Before going to the construction of the reduced functionals
$\psi^{\rm inv}$ in \equ{gsol}, let us proceed to the definition of
the kinematical Hilbert space $\HH_{\rm kin}$ and of the field operators acting in it. 
In order to have a Hilbert space we need a well defined scalar product, 
\eq
\bra{\Psi_{1}}\Psi_{2}\rangle:=\int
\mathcal{DA}\;\overline{\Psi}_{1}[A]\Psi_{2}[A]\,,
\eqn{ccccc}
with an integration measure $\mathcal{DA}$ in the space of the connections 
defined in such a way that the scalar product is compatible with the Gauss
constraint partially solved by \equ{gsol}. We therefore write
\eq
\Psi[A_{\theta}] = \Psi_{\circ}[A_\theta] \psi[A_{\theta}]\,,
\eqn{factorization}
and will look for $\psi[A_{\theta}]$. 
 The Gauss constraint takes the form \equ{reduced-func} in terms of $\psi[A_\theta]$, 
i.e., it expresses the invariance of the latter under the $\theta$-gauge transformations 
\equ{theta-transf}, which will be implemented in 
Subsection \ref{sec-gauss}.

In the spirit of LQG, 
we change the focus from the Lie algebra-valued connection 
$A_{\theta}$, which transforms 
inhomogeneously under gauge transformations accordingly to \equ{theta-transf}, 
 to the holonomies of $A_\theta$, which are elements of the gauge group. An holonomy 
$h(\g_{z,\theta_1,\theta_2})$ is 
defined over some constant $z$ path $\gamma$ $=$ $[\theta_1,\,\theta_2]$ 
in $\Sigma$ as
\begin{equation}
\label{holodef} 
h(\g_{z,\theta_1,\theta_2})=
\mathcal{P}e^{\int_{\theta_{1}}^{\theta_{2}} d\theta\;
A_{\theta}^{I}(\theta,z)\tau^{I}}\,,
\end{equation}
where $\mathcal{P}$ stands for ``path ordered product''.
This choice is motivated by the fact that under a gauge transformation of 
$A_{\theta}$, this holonomy transforms homogeneously:
\begin{equation}
\label{holotrans}
h(\g_{z,\theta_1,\theta_2})\longmapsto 
g^{-1}(z,\theta_2)\;h(\g_{z,\theta_1,\theta_2})\;g(z,\theta_1)
\end{equation}
We have to restrict to constant $z$ paths because any other paths would
involve the component $A_z$, which does not enter as an argument of
the wave functional.

Let us now define the space Cyl consisting of  all the wave functionals of the 
form \equ{factorization} with 
$\psi$ given by arbitrary finite linear
combinations of complex valued functions $f$ of the 
holonomies:
\begin{equation}
\label{cylfunc}
\psi[A_{\theta}] = \psi_{\GA,f}[A_{\theta}] 
= f\big(  h(\gamma_{z_1,\theta_1,\theta'_1}),\cdots,
h(\gamma_{z_{k},\theta_k,\theta'_k}),
\cdots,h(\gamma_{z_{N},\theta_N,\theta'_N})\big)\,,
\end{equation}
where $\GA$ denotes the ``graph'' defined as the (finite) set of paths 
\eq
\GA=\{\gamma_{z_{k},\theta_k,\theta'_k}\,,\ k=1,\cdots,N\}\,.  
\eqn{graph-def}
(See Figure \ref{kin}.)
Elements of Cyl  are called
``cylindrical functions''. A  cylindrical function
is thus a functional of $A_\theta$ which, apart from the phase factor $\Psi_\circ$, 
depends only  on the values that
its argument takes on the graph $\GA$.

\begin{figure}[ht]
\centering
{
\includegraphics[scale=0.3]{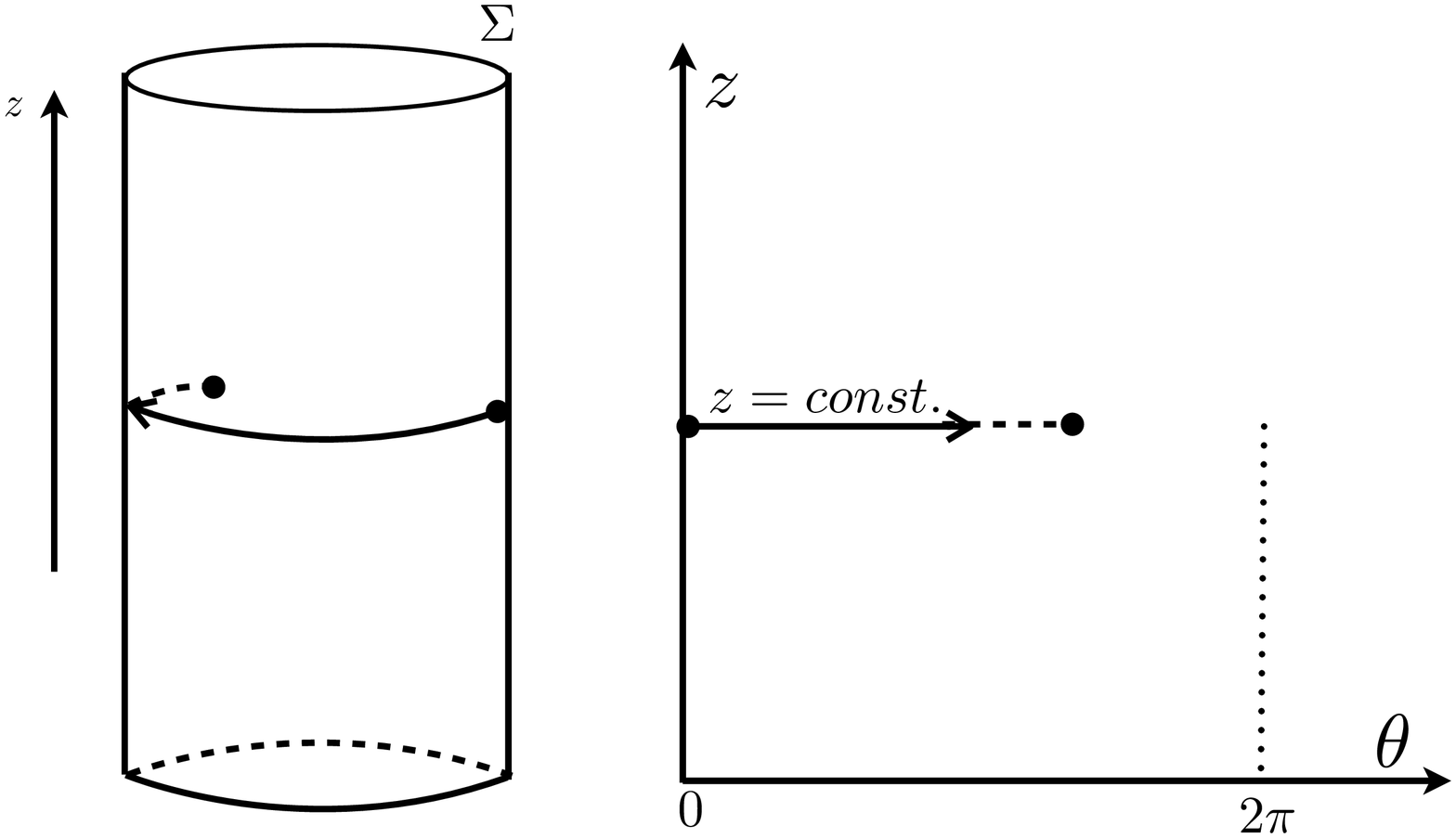}
}
{\includegraphics[scale=0.23]{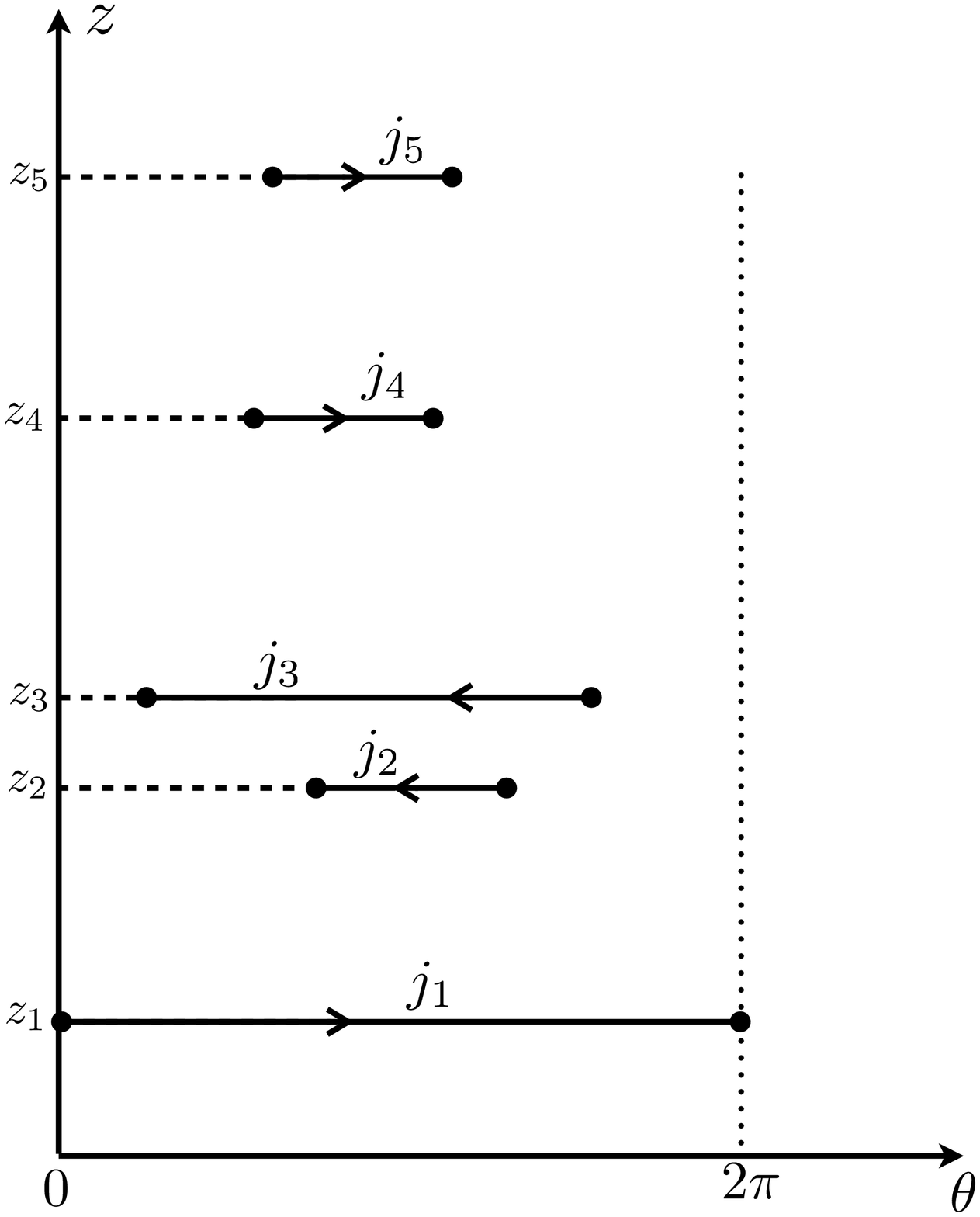}
}
\label{kin}
\caption{The construction of graphs in the space $\Sigma$, with the topology of a 
cylinder. The figure at left shows a single curve at some constant value for 
the coordinate $z$, and the figure at right is the representation of a 
particular graph in $\Sigma$, made from the disjoint union of curves at 
different ``heights'' $z_{k}$.}
\end{figure}

Since we are dealing with wave functionals of the form \equ{factorization} 
 with now $\p$ = $\p_{\GA,f}$ being a function of a finite number of holonomies,
which in turn are elements of the gauge group, we can now count on the 
invariant Haar measure of the gauge group $G$ to define a scalar product in Cyl. 
Let us first define it for two state vectors\footnote{We use
Dirac's notation, where the wave functional may be written as
$\Psi_{\GA,f}[A_\theta]$ $=$ $\vev{A_\theta|\GA,f}$.}
$\ket{\GA,f}$ and $\ket{\GA,f'}$
associated with the same graph $\GA$ \equ{graph-def}, $f$ and $f'$ being two arbitrary
integrable functions on $G^{\times N}$:
\begin{equation}\label{scalar-product}
\bra{\Gamma,f}\Gamma,f^{\prime}\rangle=
\int \prod_{k=1}^{N}dh_k\;
\overline{f(h_1\cdots h_{N})}\; f^{\prime}(h_1\cdots h_{N})\,,
\end{equation}
where $dh_{k}$ is the Haar measure used to integrate over
the group element $h_k$. 
  The Haar measure being 
normalizable for a compact group, the scalar product is well defined, 
provided the
functions $f$ on the group defining the cylindrical functions as in \equ{cylfunc}, are
integrable.  We shall take a normalized measure: 
$\int dh(h)=1$.

In the general case where the  two vectors are associated to two
distinct graphs $\GA$ and $\GA'$, we can still apply the definition
\equ{scalar-product}, but for the graph $\GA'{}'$ $=$ $\GA\cup\GA'$. Indeed, the functions 
$f(h_1\cdots h_{N})$ and $f'(h_1\cdots h_{N'})$ may be considered both 
as functions of the $N'{}'$ arguments corresponding to the paths of
$\GA'{}'$.

 We note that this scalar product is a genuine scalar product for the
vector space Cyl -- whose elements are the wave functionals
\equ{factorization} with $\p[A_\theta]$ given by \equ{cylfunc}. Indeed,
it is a positive definite sesquilinear form in Cyl.

Finally, the kinematical Hilbert space $\mathcal{H}_{\rm kin}$ is the
Cauchy completion $\overline{\rm Cyl}$ of Cyl with respect to the scalar 
product we have just defined.
\vspace{4mm}

\noindent\emph{\bf Remarks}
\begin{itemize}
\item  In contrast to theories usually considered in the loop quantization 
framework, such as $BF$ theories, gravity, etc., the wave functionals
which are 
solutions of the Gauss constraint do not depend here only on the holonomies
of the connection. Indeed, the phase factor $\Psi_\circ$ in
\equ{factorization},
being an integral of $A_{\theta}$-dependent group elements over all space, cannot be expressed
in terms of holonomies. However this does not cause any problem since 
our definition of the scalar product
is independent of this phase factor.
\item By construction, the scalar product is gauge invariant and it is
also invariant under the diffeomorphisms which preserve the polarization
choice.
\end{itemize}

\subsection{A basis for the kinematical Hilbert space}\label{basis-kin}

From the Peter-Weyl theorem~\cite{barut},  the 
wave functional corresponding to a graph
$\GA$ and to a function $f$ according to \equ{cylfunc} can be expanded 
in terms of a group representation basis as\footnote{From now on we restrict ourselves to 
the gauge group SU(2).}:
\begin{equation}\label{kinbasis}
\ba{l}
\Psi_{\GA,f}[A_\theta]  = \Psi_{\circ}[A_\theta]
f\big(  h(\gamma_{z_1,\theta_1,\theta'_1}),\cdots,h(\gamma_{z_{k},\theta_k,\theta'_k}),
\cdots,h(\gamma_{z_{N},\theta_N,\theta'_N})\big)   \\
\phantom{\Psi_{\GA,f}[A_\theta]}
= \Psi_{\circ}[A_\theta]\;
\dsum{\vec{j},\vec{\alpha},\vec{\beta}}{}
c^{\alpha_{1}\cdots \alpha_{N}}_{\beta_{1}\cdots \beta_{N}, j_{1}\cdots j_{N}}\;
R^{j_{1}, \beta_{1}}_{\alpha_{1}}(h(\gamma_{z_1,\theta_1,\theta'_1}))\cdots 
R^{j_{N}, \beta_{N}}_{\alpha_{N}}(h(\gamma_{z_{N},\theta_N,\theta'_N}))\,,
\ea
\end{equation}
where $R^{j,\beta}_{\alpha}(h)$ is a matrix element of the spin $j$ unitary 
irreducible representation of the group element $h$.
Thus, to every path $\g_{z_k,\theta_k,\theta'_k}$ of 
the graph $\GA$ we associate a spin $j_k$
representation of $G$ = SU(2). 

Moreover, the terms in the right-hand-side form 
an orthonormal system of vectors $\ket{\GA,\vec j,\vec\a,\vec\b}$:
\eq\ba{l}
\vev{\GA,\vec j,\vec\a,\vec\b\,|\,\GA,{\vec j}',{\vec\a}',{\vec\b}'} \\
= \dint \prod_{k=1}^{N}dh_k\;
\overline{R^{j_{1}, \beta_{1}}_{\alpha_{1}}(h_1)\cdots 
R^{j_{N}, \beta_{N}}_{\alpha_{N}}(h_N)}
\,R^{j'_{1}, \beta'_{1}}_{\alpha'_{1}}(h_1)\cdots 
R^{j'_{N}, \beta'_{N}}_{\alpha'_{N}}(h_N)
= \d_{\vec j{\vec j}'} \d_{\vec\a{\vec \a}'} \d_{\vec \b{\vec \b}'}\,.
\ea
\eqn{orthon}
The contribution corresponding to a graph with a spin 0 path 
is equal to the contribution corresponding to the smaller graph obtained by erasing
this path. In order to  avoid a double counting, we will only include
in the summation \equ{kinbasis}, from now on, 
terms where all the $j_k$ are different from zero. In view of this
restriction, it is clear that the set of vectors
\eq
\lac\,\ket{\GA,\vec j,\vec\a,\vec\b}\,;\, 
\forall\GA, \forall (\vec j,\vec\a,\vec\b)\,\rac
 \,\bigcup\, \lac \ket{\phantom{\vec j}\emptyset\phantom{\vec j}}\rac\,,
\eqn{kin-basis}
 where the vector $\ket{\emptyset}$ is associated to the empty graph
$\Gamma=\emptyset$, corresponding to the wave functional
$\Psi_\emptyset[A_\theta] = \vev{A_\theta|\emptyset}=1$,
is an orthonormal  basis of the kinematical Hilbert space
$\mathcal{H}_{\rm kin}$. As a consequence of the Peter-Weyl formula
\equ{orthon}, valid for any spin including spin 0, vectors
associated to different graphs are orthogonal. Thus the 
kinematical Hilbert space is a direct sum:
\eq
\mathcal{H}_{\rm kin} =
\bigoplus_{\Gamma}\mathcal{H}_{\Gamma}\,,
\eqn{ddddd}
where $\mathcal{H}_{\Gamma}$ is the Hilbert space associated with the graph
$\GA$, and the summation is made over all possible graphs. 
Whereas each $\mathcal{H}_{\Gamma}$ is a separable Hilbert space,
this is obviously not the case for $\mathcal{H}_{\rm kin}$.

 
\subsection{Solution of the Gauss constraint}\label{sec-gauss}

In the preceding section we saw that the change of the configuration 
variable from $A_{\theta}$ to the holonomies $h[A_\Gamma]$ allows for 
a well defined scalar product and consequently, a Hilbert space.
Although one cannot define a local operator $\hat A_{\theta}$ acting in
$\Hil_{\rm kin}$
 because of the discontinuity of the scalar 
product~\cite{rovelli,thiemann,ash-lew}, one can do it
for the holonomy operator $\hat{h}[A_{\theta},\g]$
associated to a path $\gamma$ at constant $z$: 
\begin{equation}
\langle A_\theta \vert \hat{h}[A_{\theta},\g] | \GA,f\rangle=
h[A_{\theta},\g]  \vev{A_\theta |  \GA,f}\,,
\end{equation}
for any basis vector $\ket{\GA,f}$. The resulting right-hand-side is indeed
an element $\vert\widetilde{\Gamma},\widetilde{f}\rangle$ of Cyl, 
associated to a new graph $\widetilde{\Gamma}$ equal to the union
$\Gamma\bigcup \{\gamma\}$ and to the function $\tilde f$ = $\hat{h}[A_{\theta},\g]f$. 

 Although the Gauss constraint -- the infinitesimal generator of gauge  
trans\-for\-ma\-tions -- expressed by \equ{reduced-func},
exists as a well defined operator\footnote{ The proof follows the one given 
by~\cite{thiemann,ash-lew,han} in a more general context. }, 
we will impose the constraint 
in the form of the invariance under all finite gauge transformations as in the standard LQG 
approach~\cite{rovelli,thiemann,ash-lew,han}. 
More precisely, taking into account the phase factor $\Psi_\circ$ in 
\equ{factorization} or \equ{kinbasis}, we will demand the gauge invariance of
the reduced wave functional $\p$ given by \equ{cylfunc} in terms of
holonomies. In view of the transformation law \equ{holotrans}, it is
clear that the gauge invariant reduced functionals are functions of the
trace of the holonomies along closed paths (cycles), i.e. of the Wilson loops 
\eq
h_z=\Tr h(\g_{z,\theta,\theta})\,,
\eqn{Wilson-loop}
which depend on the coordinate $z$, but not on the base angle $\theta$.
Thus the graphs are now sets $C$ of cycles, each cycle being characterized by its
``height'' $z$: $C$ $\leftrightarrow$ $(z_1,\cdots,z_n)$. 

This condition of gauge invariance defines the Hilbert space $\Hil_{\rm Gauss}$. 
Its basis is the orthonormal set of ``spin network'' vectors $\ket{C,J}$
which are given by the traces of the basis vectors  of $\Hil_{\rm kin}$:
\eq
\vev{A_\theta |C,J}
= \Psi_{\circ}(A_\theta) \prod_{k=1}^n\chi^{j_k}(h_{z_k})\,,\quad
\mbox{with}\  \chi^j(h_{z})= \Tr R^{j}(h_{z})   \,,
\eqn{spin-network-basis}
where $J$ stands for $(j_1,\cdots,j_n)$ and 
$R^j$ is the spin $j$ representation of SU(2). One has the
orthonormality property
\eq
\vev{C,J | C',J'} = \d_{C,C'}\d_{JJ'}\,.
\eqn{orth}
Let us define $S_0$ as the vector space of all finite linear combinations of
spin-networks,
\eq
\ket{\Psi} = 
\sum_{n=1}^{N} c_n \ket{C_n,J_n}\,.
\eqn{invfun}
The Hilbert space $\Hil_{\rm Gauss}$ is the Cauchy completion of
$S_0$. It decomposes in orthogonal subspaces in a way analogous to
$\Hil_{\rm kin}$:
\eq
\mathcal{H}_{\rm Gauss} =
\bigoplus_{C}\mathcal{H}^C_{\rm Gauss}\,.
\eqn{Gauss-direct-sum}
The Hilbert space $\mathcal{H}^C_{\rm Gauss}$ associated to a 
single
graph $C$ is separable, but $\mathcal{H}_{\rm Gauss}$ is not,
since the graphs are indexed by $n$-arrays of real numbers.


\subsection{Diffeomorphism invariance}\label{sec-diff}

We observe that $\mathcal{H}_{\rm Gauss}$ is not invariant under
the space diffeomorphisms. In particular, its basis vectors 
$\ket{C,J}$, 
depend explicitly  on the $z$ coordinates of the cycles constituting the  graph
$C$ and are therefore not invariant under changes of the coordinate $z$. (They are
however invariant under changes of the coordinate $\theta$.) This has to
be contrasted with the situation in the classical theory, where the
fulfilment of the Gauss constraint automatically ensures full space
diffeomorphism invariance (see \equ{diff-A}). 
We note from \equ{diff-A} that a diffeomorphism along the $z$ 
coordinate generated by a vector field $\xi=(\xi^z,0)$, acts on the
configuration variable $A_\theta$ as $\LL_\xi A_\theta$ $\approx$ $D_\theta(\xi^z A_z)$, 
i.e. as a gauge transformation with parameter $\xi^z A_z$. However, when
applied to a wave functional, $A_z$ must be replaced by the operator
defined in \equ{ope1}. Such a ``gauge transformation'' was not
contemplated when we solved the Gauss constraint in Subsection \ref{sec-gauss}.
Therefore, we have still to implement this part of diffeomorphism
invariance, namely invariance under the $z$-diffeomorphisms 
\eq
z'=z'(z)\,,\quad\theta'=\theta\,.
\eqn{z-diff}
 The more general diffeomorphisms generated by vectors
$\xi=(\xi^z,\xi^\theta)$ -- which modify the polarization -- are left
aside for the time being.

In the same way as we have proceeded with the Gauss constraint, we will
impose invariance under the \emph{finite} $z$-diffeomorphisms \equ{z-diff} due to the
difficulty of defining their infinitesimal generator.

We will follow the standard  group averaging 
method~\cite{rovelli,thiemann,ash-lew}, based here on the Gel'fand
triple~\cite{gelfand-triple} $S_0$ $\subset$ $\mathcal{H}_{\rm Gauss}$
$\subset S_0'$, where $S_0$ is the subspace of finite linear combinations 
of spin-networks defined at the end of the
preceding subsection, dense in $\mathcal{H}_{\rm Gauss}$, and $S_0'$
its dual, whose
elements are the complex valued linear functionals $\F$ of $S_0$:
\eq
\F: S_0\ \to \ \complex\,,\quad 
\Psi\ \mapsto \ \vev{\F,\,\Psi}\in \complex\,, 
\eqn{linear-fnal}
where we  use Schwartz notation $\vev{\ ,\ }$ for
functionals~\cite{schwartz}. The scalar product  \equ{scalar-product} 
in $\HH_{\rm Gauss}$
being explicitly invariant under  all space diffeomorphisms, any such 
diffeomorphism $\f$ is represented by a unitary operator $U_\f$. The
action of $\f$ in $S_0'$ is then defined by duality:
\eq
\vev{U_\f\F,\,\Psi} = \vev{\F,\,U_{\f^{-1}}\Psi}\,.
\eqn{action-diff}
We will concentrate on the  $z$-diffeomorphisms \equ{z-diff}. 

The $z$-diffeomorphism invariant states are now given by vectors of $S_0'$
constructed from any vector $\ket{\Psi}$ of $S_0$ by applying to it the
operator $P_{\rm Diff}$ -- a functional ``projector'':
\eq
P_{\rm Diff}: S_0 \to S_0'\,,
\vev{P_{\rm Diff}\Psi,\,\Psi'} = \dsum{\Psi''}{} \vev{\Psi''|\Psi'}\,,\
\forall\ \ket{\Psi'}\in S_0\,,
\eqn{projector}
where the sum is done over all the vectors $\ket{\Psi''}$ of $S_0$ which may be
obtained from $\ket{\Psi}$ by a $z$-diffeomorphism: $\ket{\Psi''}$ $=$
$U_\f\ket{\Psi}$.
 The sum in \equ{projector} is always finite. Indeed, 
the vectors $\ket{\Psi}$ and $\ket{\Psi''}$ are both finite
superpositions of spin-networks vectors -- see \equ{invfun}: 
\eq
\ket{\Psi}= \dsum{m=1}{M} c_m \ket{C_m,{J}_m}\,,\qquad
\ket{\Psi''} = \dsum{n=1}{N} c''_n \ket{C''_n,{J}''_n}\,.
\eqn{eeeeeee}
Since two spin-network vectors
are orthogonal if their respective graphs are different, the summation 
in \equ{projector} is
restricted to those $\ket{\Psi''}$ which have at least one graph $C''_n$
in its expansion
which coincides with a graph $C_m$ of the expansion of $\ket{\Psi}$. 
But, in the present theory where space is one-dimensional, given a spin-network vector
$\ket{C_m,{J}_m}$, there are only two such
spin-network vectors: $\ket{C_m,{J}_m}$ itself and the vector
$\ket{C''_n,{J}''_n}$, with $C''_n$, 
 obtained from the first one by inverting the order of the ``edges'' of the 
graph $C_m$ , i.e.  
of the coordinates $z_k$ which label these edges -- with  the
restriction that the spins ${j}''_k$ attributed to the edges still match after this
inversion.

The functionals $P_{\rm Diff}\Psi$ defined in this way are $z$-diffeomorphism invariant,
as a consequence of the equation \equ{action-diff} 
which defines the way elements of $S_0'$ transform.
They span by definition the vector space $\mathcal{H}_{\rm Phys}$ $\subset$ $S_0'$,
which we will show to be the physical Hilbert space of the theory. 

We observe that the only element of this space which is also an element 
of $\mathcal{H}_{\rm Gauss}$ is the trivial 
state $\ket{0}$ -- the ``vacuum'' -- defined by 
\eq
\vev{A_\theta | 0}= \Psi_{\circ}(A_\theta) \,,
\eqn{vacuum}
where $\Psi_{\circ}$ is the phase factor \equ{Psi_0}, which is
obviously diffeomorphism invariant, in particular under the 
$z$-diffeomorphisms considered here.

We define now the scalar product in $\mathcal{H}_{\rm Phys}$  by
\begin{equation}
\label{prodintdiff}
\bra{P_{\rm Diff}\Psi_{1}}P_{\rm Diff}\Psi_{2}\rangle 
:= \langle P_{\rm Diff}\Psi_{1},\Psi_{2}\rangle
\end{equation}
where $\ket{\Psi_{1}}, \ket{\Psi_{2}} \; \in S_0$.
This product is independent of the 
particular state $\ket{\Psi_{2}}$ we used to define 
$\ket{P_{\rm Diff}\Psi_{2}}$.  

By construction (see \equ{projector}), vectors of $\mathcal{H}_{\rm phys}$  
only depend on the equivalence classes of vectors of $S_0$ under $z$-diffeomorphisms.
In particular, a vector defined by \equ{projector} from a spin-network 
$\ket{C,J}$ does not depend on the particular positions $z_k$ of the cycles 
constituting the graph $C$, but only on the number of such cycles -- and of the 
spin values associated to each of them. Following the LQG terminology~\cite{rovelli}, 
let us call such a vector an s-knot and denote it by $\ket{j_1,\cdots,j_N}$ $\equiv$ 
$\ket{J}$:
\eq
\ket{J}=P_{\rm Diff} \ket{C,J}
\eqn{s-knot}

The scalar product of two s-knots is given by
\eq
\vev{J|J'} 
= \vev{j_{1},\cdots,j_{N} | j'_{1},\cdots,j'_{N'}} 
=\delta_{NN'}\lp\delta_{j_{1}j_{1}'}\cdots \delta_{j_{N}j_{N'}}
 +\delta_{j_{N}j'_{1}}\cdots \delta_{j_{1}j'_{N'}}\rp\,.
\eqn{phys-scal-prod}
The second term in the right-hand side is due to the existence, mentioned above, 
of $z$-diffeomorphisms which preserve a graph but reverse the order of its
cycles. Thus,  provided one identifies a vector
$\ket{j_1,j_2,\cdots,j_N}$ with its ``reversed''
$\ket{j_N,\cdots,j_2,j_1}$,
the s-knot states 
provide an orthonormal basis of $\mathcal{H}_{\rm Phys}$.

 The vectors  \equ{s-knot} being completely characterized by finite
sets of half-integer numbers $J=\{j_1,j_2,\cdots,j_N\}$, 
are clearly invariant under {\it all diffeomorphisms}, beyond being
solutions of the Gauss constraint. In particular, they do not depend on
the choice of the polarization.
$\mathcal{H}_{\rm Phys}$ is thus the
physical state space of the Chern-Simons theory on a cylinder. The set of
s-knots vectors being countable, this Hilbert space is separable.

\section{Conclusions}\label{conclusions}
We found in Chern-Simons theory a great opportunity to discuss the main ideas of 
loop quantum gravity, concerning the method by itself. In fact, this theory fits 
exactly in the framework which was used to think on  quantum gravity in the 
beginning~\cite{rovelli-smolin}, where the quantum states were generated
by Wilson loops of the Ashtekar connection~\cite{ashtekar}, 
the so called ``loop states''. 

What we have concretely done here is a continuation of~\cite{djt}. 
In this reference the authors showed the path to get a physical quantum state 
based on the implementation of the Gauss constraint, 
which could not be implemented in the usual way because of 
the difficulty to define the quantum operators based on the choice of a
Schr\"odinger representation. In order to really define a Hilbert space,
we needed, after having fixed a particular topology (as suggested in~\cite{djt}) 
for the spatial slice $\Sigma$, to change the configuration variables 
from the connection to its holonomies, following the prescriptions of
Loop Quantum Gravity. The result, for the chosen topology, is a physical
Hilbert space whose basis is indexed by a nonnegative integer $N$ and, for each $N$, 
by an ordered set of $N$ half integer numbers $j_1$, $\cdots$, $j_N$ -- the spins.

Our choice of $A_\theta$ -- the connection component in the direction of
the compact space dimension -- as the configuration variable, and of
$A_z$ -- the component in the noncompact direction -- as the conjugate
variable, induced the breaking of general covariance. This is the reason
why the Gauss constraint was not sufficient to assure full
diffeomorphism invariance, which was finally recovered by imposing the
invariance under the diffeomorphisms along the $z$ coordinate.

 One could wonder on our choice \equ{factorization}, with
\equ{cylfunc}, for the elements of the vector space Cyl whose Cauchy
completion yields the kinematical Hilbert space $\mathcal{H}_{\rm kin}$.
The main reason for doing so is that it allows for a well defined scalar 
product and for a simple solution of
the Gauss constraint in terms of $\theta$-gauge invariant functions of the 
$z$ = constant holonomies. Another approach\footnote{We thank a referee for suggesting it.},
which would apparently be more 
in the spirit of loop quantum gravity, would consist in taking, as a
kinematical space, the functions of the $z$ = constant holonomies,
without the phase factor $\Psi_\circ$, then trying to construct the
Gauss constraint operator in this space and finally taking its kernel. Even if
manageable, this way would certainly be much more cumbersome than
the one we have followed.

We have taken SU(2) as the gauge group. Generalization to other compact
Lie groups looks straightforward.
The consideration of more general topologies, as well as the
construction of physical observables along the same lines are left
for future works.

\noindent{\bf Acknowledgments} We thank Laurent Freidel,  Camillo Imbimbo 
and Galen Sotkov for most valuable informations and discussions.
This work was supported
   in part by the Conselho Nacional
   de Desenvolvimento Cient\'{\i}fico e
   Tecnol\'{o}gico -- CNPq (Brazil) and 
by the PRONEX project No. 35885149/2006 from FAPES -- CNPq (Brazil).


\end{document}